\documentclass[conference]{IEEEtran}
\IEEEoverridecommandlockouts

\usepackage{cite}
\usepackage{amsmath,amssymb,amsfonts}
\usepackage{algorithmic}
\usepackage{graphicx}
\usepackage{textcomp}
\usepackage{url}
\usepackage{xcolor}
\usepackage{array}
\usepackage{booktabs}
\def\BibTeX{{\rm B\kern-.05em{\sc i\kern-.025em b}\kern-.08em
    T\kern-.1667em\lower.7ex\hbox{E}\kern-.125emX}}

\newcommand{\review}[1]{{{#1}}}

\begin{document}

\title{A Microbenchmark Framework for Performance Evaluation of OpenMP Target Offloading
} 


\author{
\IEEEauthorblockN{
    Mohammad Atif\IEEEauthorrefmark{1},
    Tianle Wang\IEEEauthorrefmark{1},
    Zhihua Dong\IEEEauthorrefmark{1},
    Charles Leggett\IEEEauthorrefmark{2},
    Meifeng Lin\IEEEauthorrefmark{1}
    }
       
    \IEEEauthorblockA{\IEEEauthorrefmark{1}Brookhaven National Laboratory, NY, USA}
    \IEEEauthorblockA{\IEEEauthorrefmark{2}Lawrence Berkeley National Laboratory, CA, USA} 
}

\maketitle
 
\begin{abstract}
We present a framework based on Catch2 to evaluate performance of OpenMP's target offload model via micro-benchmarks.
The compilers supporting OpenMP's target offload model for heterogeneous architectures are currently undergoing rapid development.
These developments influence performance of various complex applications in different ways.
This framework can be employed to track the impact of compiler upgrades and compare their performance with the native programming models.
We use the framework to benchmark performance of a few commonly used operations on leadership class supercomputers such as Perlmutter at National Energy Research Scientific Computing (NERSC) Center and Frontier at Oak Ridge Leadership Computing Facility (OLCF).
Such a framework will be useful for compiler developers to gain insights into the overall impact of many small changes, as well as for users to decide which compilers and versions are expected to yield best performance for their applications.
\end{abstract}


\section{Introduction}

In the past decade, the importance of portable parallel programming has increased due to the variety of accelerators in high-performance computing (HPC) systems.
Most large scale supercomputers today are heterogeneous and employ Graphics Processing Units (GPUs) from various vendors like Intel, AMD, and NVIDIA, each of which has its own native application programming interface (API). 
In order to use the GPUs, one needs to either port compute-intensive parts of the application to a vendor-specific API, or write the application with a portable programming model. 
Several portable programming models have been developed in the last decade as a solution to avoid code duplication and diversion for different GPU backends. Examples of such portable programming models include Kokkos~\cite{kokkos}, SYCL~\cite{sycl}, RAJA~\cite{raja}, Alpaka~\cite{alpaka}, OpenACC~\cite{openacc} and OpenMP~\cite{openmp}. 
This work focuses on OpenMP's target offload model, which has undergone rapid development in the past few years and gained increasing vendor support~\cite{parco,aaroncicd}.
OpenMP's directive-based GPU offloading strategy is also a strong candidate for performance portability of Fortran codes \cite{hsu2018performance}, although we focus our attention to C/C\texttt{++} in the following discussion. 

The quantification and reporting of HPC technologies and their evolution have traditionally been based on performance benchmarks \cite{radulovic2018hpc,hoefler2015scientific}.
Perhaps, the most famous benchmarks are the STREAM and High Performance Linpack (HPL) benchmarks \cite{stream,hpl}. 
There are collections of several excellent benchmark suites \cite{nvidia-hpc-benchmarks,araujo2023parallel,spec-accel,hpcg} that measure the run times of real-world and synthetic applications.
Most of these benchmarks calculate the entire performance of a reasonably large application with several active components during various phases of execution \cite{sayeed2008measuring}.
To narrow down performance bottlenecks, micro-benchmarking frameworks have been developed concurrently with testing frameworks such as Google Benchmark, Nonius, and Catch2 \cite{google-bench,nonius,catch2}.
Such micro-benchmarking frameworks execute a small snippet of code multiple times followed by statistical analyses to generate highly reliable metrics of computational performance.
We demonstrate that micro-benchmarks can be useful tools for comparing programming models too.

In this work, we present a framework based on \verb|Catch2| \cite{catch2} for performance evaluation of OpenMP's \verb|target offload| model via microbenchmarks.
\review{The primary motivation is to develop the necessary tools for the better evaluation of specific operations in OpenMP target offload.
The performance of the OpenMP target offload currently varies depending on the target backend and the application, as reported in ~\cite{parco,iwomp21-1,iwomp21-2}. 
We have encountered several performance issues~\cite{lin2023portable} related to specialized, operations, such as atomic, memset or scan, when porting application codes to OpenMP.
The key idea behind developing a micro-benchmarking framework is to generate high quality measurements by creating samples in contrast to running the same kernel a number of times in a loop and reporting the mean.
As a demonstration, in this paper, we use the framework to benchmark the performance of several widely used operations on NERSC Perlmutter and OLCF Frontier (see \cite{perlmutter,frontier} for architectures) and two servers with Nvidia V100 and A6000 housed at Brookhaven National Lab.

There are several advantages to benchmarking with Catch2, such as:
\begin{itemize}
    \item Easy integration: While there are several similarities between Catch2's benchmarks and Google Benchmarks in their sampling and iteration count evaluation, Catch2 is light-weight and enables a more streamlined integration of unit-testing and benchmarking framework. 
    \item Dynamic estimation of iteration count: Catch2's micro-benchmarks create samples by accounting for the clock-resolution and dynamically estimating the iteration count of the kernel by estimating its runtime. Each sample can consist of more than one run of the kernel if the available clock lacks sufficient resolution.
    \item Statistical bootstrapping: The statistical bootstrapping method first creates a set of samples and then creates several subsets of resamples to calculate the mean and standard deviation.
    The resampling technique reduces the impact of artifacts of measurement. This feature is quite important for our framework as it minimizes the chance of a fringe case biasing the measurements. It calculates confidence intervals for each run thus improving robustness and enhancing comparability across hardware, data types, threads per block, and programming models.
    \item Catch2 also ensures that the compiler does not optimize away the kernel of interest in the case its output is not being used later in the program.
\end{itemize}
The above features offer additional information as compared to standard tools such as Nsight Systems which instrument the code thus incurring profiler's overheads.
Indeed, the micro-benchmark results need to be combined with traces and profiles along with assembly to analyze the root causes of performance disparity.
However, the idea behind this framework is to pinpoint the location of good/bad performance in the wide space of compilers, programming models, compiler flags, hardware, datatypes and array sizes for a few commonly used kernels. 
Therefore, the root cause analyses are kept out of scope of the present manuscript.
The remainder of the paper is organized as follows: in Sec. \ref{sec:openmp}, we discuss OpenMP's target offload model. 
In Sec. \ref{sec:catch2}, we introduce our framework and discuss benchmarking with Catch2. 
In Sec. \ref{sec:results} we use the present framework to benchmark a few commonly used kernels and compare them with the native programming models across compiler version, hardware, data types, and compiler flags.
Finally, in Sec. \ref{sec:outlook} we summarize our findings and discuss avenues for further research.

}

\begin{figure*}
\centering
\includegraphics[width=0.99\textwidth]{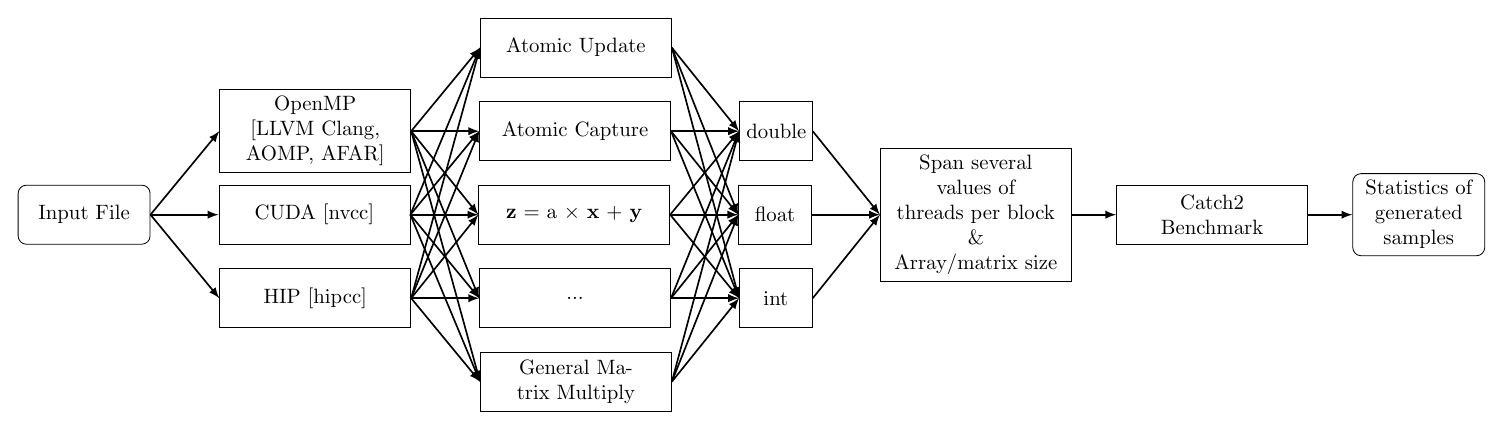}
\caption{A sketch of Catch2 based microbenchmarking suite's workflow.}
\label{fig:sketch}
\end{figure*}  

\section{Related Work}
OpenMP micro-benchmarks~\cite{epcc} were developed by the Edinburgh Parallel Computing Centre (EPCC) to measure the overhead of synchronization, loop scheduling, array operations, and task scheduling. 
However, it does not yet support the OpenMP target offload constructs. 
The SOLLVE project of the US Exascale Computing Project (ECP) has developed a validation and verification suite~\cite{sollve-vv} to test the correctness and conformance to the OpenMP specification of different compiler implementations. 
However, performance evaluation is not one of SOLLVE's main goals. 
The SPEC HPC~\cite{spec-hpc} and SPEC ACCEL~\cite{spec-accel} benchmarks evaluate the performance of different architectures and different parallel APIs using mini-applications from different scientific domains, which are much more complex than the micro-benchmarks we present here. 
The FAROS framework~\cite{faros} developed by Georgakoudis \textit{et al.} has some similarity to our framework in that it is also extensible and provides a test harness to ease the performance evaluation process.  
But it targets third-party applications that can benefit from the evaluation of different compiler and runtime options, and does not include benchmarks or microbenchmarks of its own.
While these benchmarks and testing suites provide a good overview of the current compiler support for OpenMP target offload, they lack the granularity to evaluate the performance of specific kernels. 
Randomized differential testing was employed by Laguna \textit {et al.} \cite{laguna2024testing} to test the correctness and performance anomalies of GCC, Clang, and Intel implementations of OpenMP.
However, they have so far only targeted multicore CPUs.
The BabelStream \cite{babel} benchmark measures the performance of memory transfer to/from GPU memory but does not focus on floating point operations. 
The present framework complements BabelStream by measuring only the performance of specialized operations via microbenchmarks.

\review{
HeCBench is a comprehensive suite of benchmarks that includes several microbenchmarks as well as application-level benchmarks tailored for heterogeneous computing systems \cite{hecbench}.
HeCBench includes CUDA, HIP, SYCL, and OpenMP for the test cases.
The collective benchmarks in HeCBench measure critical low‐level operations such as memory latency, kernel launch overhead, and atomic operations by calculating mean of a fixed number of executions.
The proposed framework is different from HeCBench in that (i) it creates samples and then resamples to calculate confidence intervals, and (ii) the number of executions of each kernel is a function of its execution time and is dynamically calculated.
Nevertheless, the comprehensiveness of tests in HeCBench is valuable and will inform our future development while also providing important baselines for a quantitative comparison.
CudaMicroBench~\cite{cudabench} is a collection of fourteen micro‐benchmarks specifically designed to assist CUDA performance programming. 
It demonstrated performance gaps in various CUDA kernels and formulates techniques to optimize.
It provides detailed information about coalesced memory accesses, data shuffling between threads, dynamic parallelism, etc that can help users optimize the CUDA program for performance.
It evaluates aspects such as kernel launch overhead, memory access patterns, and the effectiveness of optimization techniques (e.g., dynamic parallelism) on NVIDIA GPUs. 
Although CudaMicroBench focuses on CUDA, its methodology for isolating and analyzing low‐level performance characteristics provides excellent design principles that are indispensable to optimizing OpenMP target offloading.
In addition to the aforementioned micro‐benchmark suites, several profilers for GPU codes -- such as NVIDIA’s Nsight Systems, Nsight Compute, AMD's uProf, rocProf, Omnitrace, and TAU -- have become indispensable tools for detailed insights into kernel execution, memory transfer patterns, and other runtime behaviors.

}


\section{\label{sec:omptarget}OpenMP Target Offloading} \label{sec:openmp}

OpenMP is a compiler directive-based programming model for shared memory parallelization.
Its fork-join model has historically been used for multithreading on multicore architectures. 
OpenMP 4.0 and above extend support for parallel execution on heterogeneous architectures via the \verb|target offload| model.
\review{
In this model, the offload regions need to be annotated with} \verb|#pragma omp target| \review{directive which instructs the compiler to generate device specific code.
The data accessed by the device within the target region should either be explicitly mapped to the device using clauses like}
 \verb|map(in:)|, \verb|map(out:)|, or \verb|map(tofrom:)| \review{or allocated on the device using} \verb|omp_target_alloc| 
\review{and the corresponding pointer marked as } \verb|is_device_pointer|.
\review{OpenMP allows the developers to exploit parallelism without directly managing the low-level details of the GPU architecture by splitting the work into teams and threads.
The OpenMP runtime creates a number of teams that are analogous to a CUDA's blocks and allows multiple teams to execute concurrently whenever sufficient resources are available.
Each team with a number of threads is scheduled independently on the GPU which the threads execute the code in parallel.
}
Among the compilers supporting \verb|target offload|, LLVM Clang and GCC are community-developed, whereas NVIDIA's nvc\verb|++|, AMD's amdclang, AOMP, AFAR, and Intel's icpx are vendor-developed. 
As most vendor compilers are derivatives of LLVM Clang, in this work we will focus on LLVM Clang's performance. 
However, the framework developed here can easily be extended to any compiler provided appropriate compiler flags are included.

The compiler-directive based OpenMP \verb|target offload| has distinct advantages for GPU offloading.
It is easy to implement with incremental porting of a serial code and does not require major changes when porting from an existing C\verb|++| code.
It permits simultaneous device and host parallelization and is interoperable with other programming models.
It also has good integration with build systems like \verb|CMake| and \verb|make|.
Many specialized operations (e.g. \verb|atomic|) yield lower performance than native programming models, whereas some operations like \verb|scan| and \verb|memset| are not supported on GPUs. 
Lastly, it is also important to tune the number of threads per team and use suitable compiler flags as they sometimes drastically improve the performance \cite{atif2023evaluating,lin2023portable}. 
Some of the compiler flags that may improve the application performance with LLVM Clang include 
\begin{verbatim}
-fopenmp-cuda-mode,
-foffload-lto,
-fopenmp-assume-no-threadstate,
\end{verbatim}
and since Clang-16 
\begin{verbatim}
-fopenmp-assume-no-nested-parallelism.
\end{verbatim}

LLVM Clang also offers features such as optimization remarks and debug environments.
The optimization remarks, invoked with flags 
\begin{verbatim}
-Rpass=openmp-opt, 
-Rpass-analysis=openmp-opt, 
-Rpass-missed=openmp-opt,
\end{verbatim}
provide details about the location or movement of data and insights into performance improvement or degradation.
The available debugging options are the flag \verb|-fopenmp-target-debug=<N>| or the environment variable \verb|LIBOMPTARGET_DEVICE_RTL_DEBUG=<N>| for compile time and runtime debugging respectively.
The environment variable \verb|LIBOMPTARGET_INFO=-1| also provides detailed runtime information related to offloaded kernels.

Evaluating the performance of portable programming models is one of the main goals of High Energy Physics -- Center for Computational Excellence (HEP-CCE) \cite{atif2023evaluating}.
We have encountered wide disparity in performance when comparing OpenMP target offload with other programming models for FastCaloSim \cite{atif2024porting} and WireCellGen \cite{lin2023portable} testbeds.
For example, (a) FastCaloSim's `simulate' kernel with OpenMP target offload is slower than CUDA on Nvidia A100, whereas, it is faster than HIP on AMD Mi100, and (b) LLVM Clang's cuda-mode flag drastically improves run times on Nvidia GPUs but makes marginal impact on AMD GPUs \cite{atif2024porting}.
Such observations lead to questions about performance hotspots, runtime environments, software versions, and hardware.
The present framework is a product of analyzing HEP-CCE testbeds with the aim to demonstrate performance disparity in an environment devoid of complex dependencies and will thus help in answering the above questions.


\section{Microbenchmarking with Catch2}
\label{sec:catch2}

\begin{figure}
\centering
\includegraphics[width=0.49\textwidth]{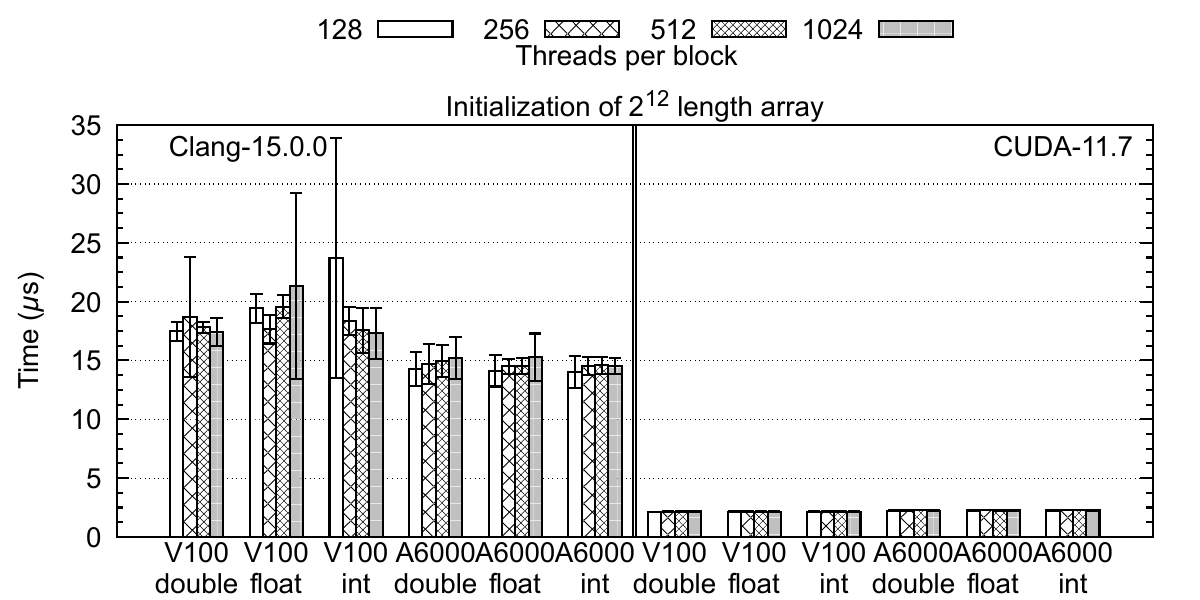}
\\
\includegraphics[width=0.49\textwidth]
{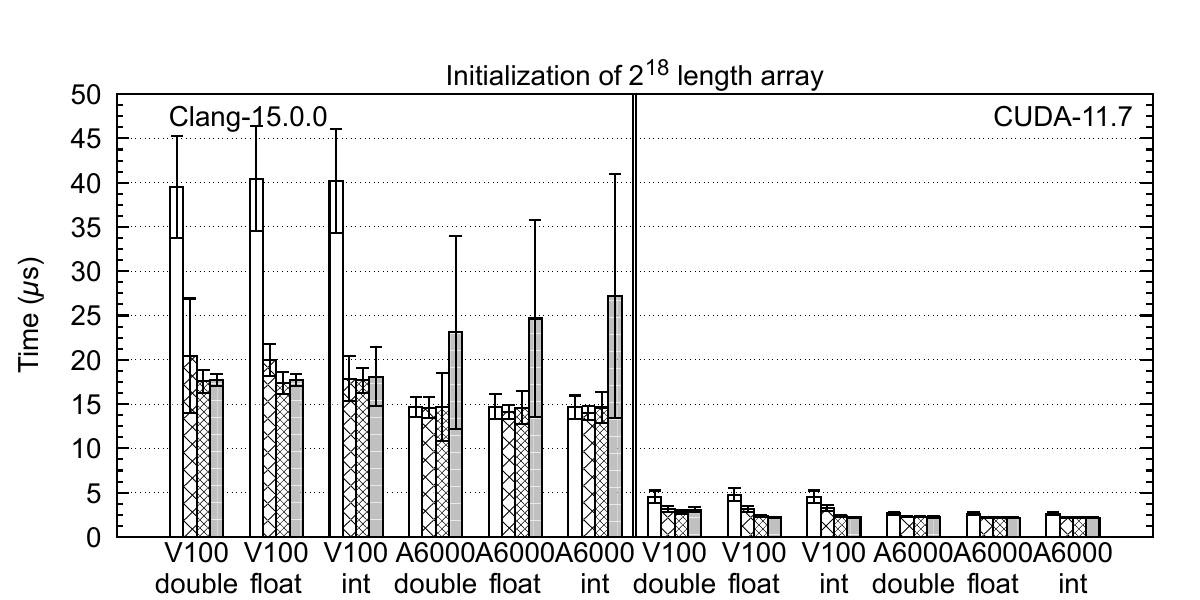}
\caption{Comparing array initialization time on Nvidia V100 and A6000: Clang-15 vs CUDA-11.7 across datatypes and threads per block.}
\label{fig:array_init_nvidia_a6000_v100}
\end{figure} 

\begin{figure}
\centering
\includegraphics[width=0.49\textwidth]{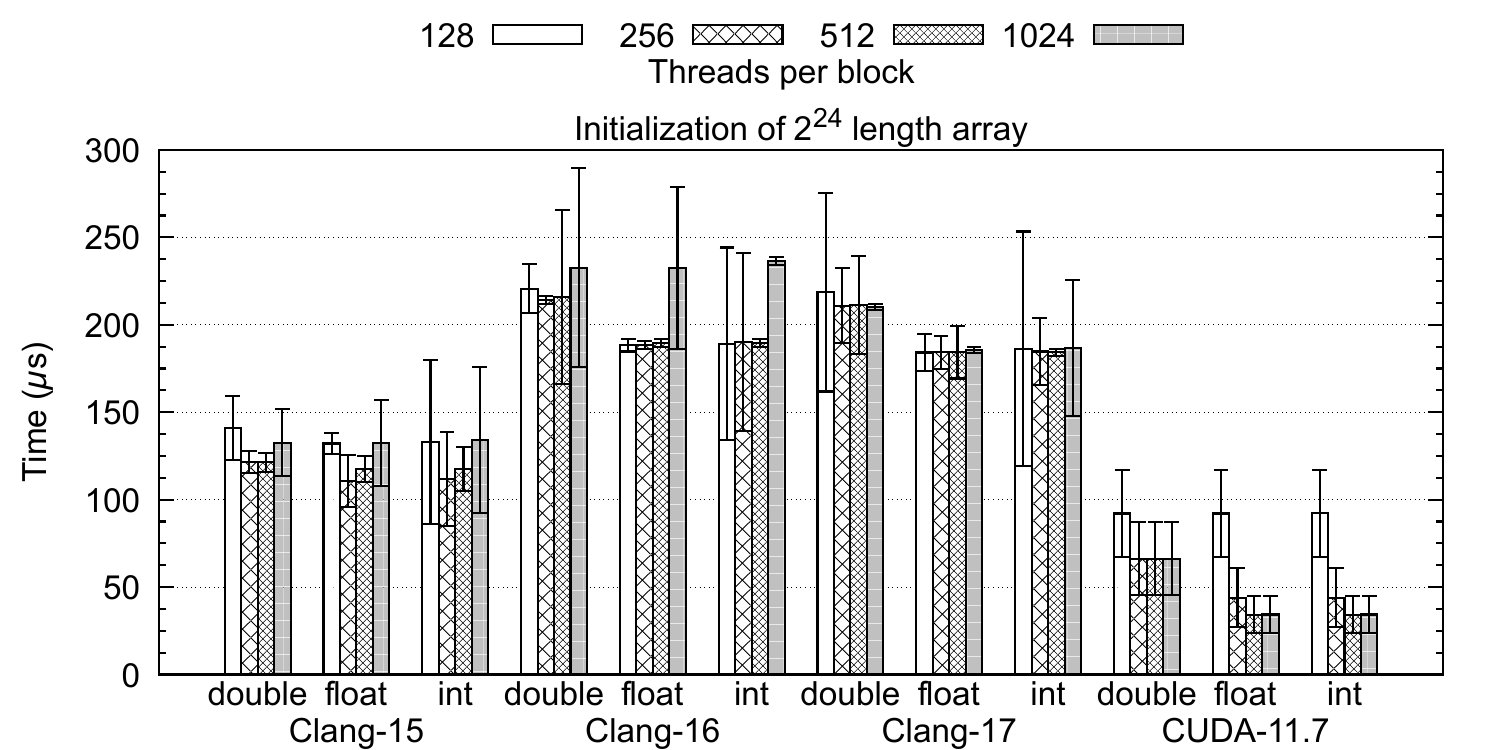}
\caption{Comparing array initialization time on Perlmutter: varying threads per block and datatype.}
\label{fig:array_init_perl}
\end{figure}

The compilers supporting OpenMP's \verb|target offload| model are currently undergoing rapid development.
These developments influence performance of various applications in different ways.
Traditionally, performance optimization in the context of scientific computing requires finding bottlenecks and improving/accelerating their behavior.
Widely different physics-based simulation suites often encounter similar bottlenecks in the form of operations like reduction, communication, sorting, etc.
Thus, evaluating performance of such operations in individual microbenchmarks and accelerating them has the potential to affect many applications at once. 
Hence there is a need for a framework to quantify the effect of these developments via micro-benchmarks.
Such a framework will be useful for compiler developers to gain insights into the overall impact of many small changes, as well as for users to decide which compilers and versions are expected to yield best performance for their applications.

Catch2 is primarily a unit-testing framework for C\verb|++| applications which has incorporated microbenchmarking to its wide list of features.
It owes its popularity to being a lightweight header only framework till version 2.x.
It is quite straight-forward to embed in complex applications and compiles easily.
Catch2 comes with two benchmarking macros, namely  \verb|BENCHMARK| and \verb|BENCHMARK_ADVANCED|, that expand to lambda expressions.
Catch2's benchmarking macros calculate the number of runs in each sample depending on their execution time.
The final statistics are calculated with an option for statistical bootstrapping.
There are several command line options such as
\begin{verbatim}
--benchmark-samples,
--benchmark-resamples,
--benchmark-confidence-interval,
--benchmark-warmup-time,
\end{verbatim}
to quantify statistics of benchmarks, and \verb|--input-file| to select execution of a subset of benchmarks. 
Each command line option has a default value and can also be set by the user at the runtime.
Figure \ref{fig:sketch} depicts the sketch of Catch2 based microbenchmarking suite's workflow.

The syntax for using \verb|BENCHMARK| macro is as follows:
\begin{verbatim}
BENCHMARK("initialize array") { 
  return function_name (device_array_ptr, 
  array_size, block_size, num_blocks); 
};    
\end{verbatim}

The syntax for measuring the execution time of more complex kernels, such as \verb|zaxpy| (see Sec. \ref{sec:zaxpy}) using \verb|BENCHMARK_ADVANCED| is as follows:
\begin{verbatim}
allocate_x_y_z_host ();
allocate_x_y_z_device ();
BENCHMARK_ADVANCED("zaxpy")(
    Catch::Benchmark::Chronometer meter) {
    
    initialize_x_y_z_host ();
    
    copy_x_y_z_host_to_device ();
    
    meter.measure([x, y, z, a, array_size, 
    block_size, num_blocks] { return 
    zaxpy(x, y, z, a, array_size, 
    block_size, num_blocks); } 
  );
  copy_z_device_to_host ();
};
\end{verbatim}
where the zaxpy kernel for CUDA/HIP is
\begin{verbatim}
template<typename T>
__global__ void saxpy ( T* result_dev, 
    T* data_x_dev, T* data_y_dev, 
    const T fact, const int size ) {

    int i = blockIdx.x * blockDim.x 
                      + threadIdx.x;
  
    if(i < size) {
        result_dev[i] = data_y_dev[i] 
            + fact * data_x_dev[i];
    }
}
\end{verbatim}
whereas for OpenMP it is 
\begin{verbatim}
template <typename T>
void saxpy ( T* result, T* data_x_device, 
          T* data_y_device, const T fact, 
          const int N, const int nblocks, 
                  const int blocksize ) {

  #pragma omp target is_device_ptr(result, 
             data_y_device, data_x_device)
  #pragma omp teams distribute parallel for 
  num_teams(nblocks) num_threads(blocksize)
  for(int i=0; i<N; i++) {
    result[i] = data_y_device[i] 
       + fact * data_x_device[i];
  }
}    
\end{verbatim}
In the above example, only the time for the \verb|zaxpy| kernel's launch and execution is measured. The other functions within the benchmark, although repeated for each run, are not included in time measurement. 

To illustrate the statistics one could obtain from this framework we analyse the run times of a kernel that simply initializes an array with zeros.
The memory is allocated on the \review{device} using \verb|omp_target_alloc|, thus the benchmark does not measure memory allocation or host to device data transfer time.
We see from Fig. \ref{fig:array_init_nvidia_a6000_v100} that Clang-15 is considerably slower than CUDA-11.7 for array sizes of $2^{12}$ and $2^{18}$ for this basic operation on Nvidia V100 and A6000 across datatypes and threads per block.
\review{It should be noted that when varying the number of threads per block the total number of teams is also modified accordingly.}
We also notice that Clang has a more pronounced standard deviation than CUDA.
On Nvidia V100 for array length $2^{18}$ we observe that 128 threads per block are almost twice as slow than 256, 512, 1024.
Figure \ref{fig:array_init_perl} plots the statistics obtained from a kernel initializing an array of different data types and size $2^{24}$. 
\review{It is observed that the overall performance of LLVM Clang has deteriorated when updating from 15, however, the performance of Clang-17 has more similarity with CUDA in that floats and ints are faster than doubles.}

\begin{table*}[]
    \centering
    \caption{GFLOPs/s obtained from the present Catch2 based framework (100 samples) compared with std::chrono timings (averaged over 100 kernel executions). The [S/D]GEMM kernel is $\alpha * \cal{A} \times \cal {B} + \beta * \cal{C}$ where $\cal{A}, \cal {B}, \cal{C}$ are N x N matrix of random numbers, S and D denote single and double precision respectively, and $\alpha = 1, \beta = 0.5$. }
    \begin{tabular}{c|c|c|c|c|c }
    Kernel (N) & Catch2 (mean) & Catch2 (max) & Catch2 (min) & std::chrono (mean of 100) & \% deviation \\
    \hline \\
    SGEMM \\
    1024 & 9596.88  & 10026.91 & 9443.65  &	9588.39	 &	0.088 \%	\\
    2048 & 18095.86 & 18447.49 & 18042.28 &	18086.81 &	0.050 \%	\\
    4096 & 24171.47 & 24274.66 & 24130.21 &	24166.77 &	0.019 \%	\\
    8192 & 25692.29 & 25707.72 & 25258.26 &	25694.16 &	-0.007 \%	\\			
    \hline
    DGEMM \\
    1024 & 441.44   & 441.48   & 440.98   & 441.11	 &  0.075 \%	\\	
    2048 & 504.55   & 504.56   & 504.52	  &	504.55	 &	0.001 \%	\\	
    4096 & 546.97	& 546.98   & 546.20	  &  546.96	 &	0.003 \%	\\
    8192 & 542.17	& 542.17   & 540.76	  &	542.13	 &	0.006 \%	\\	    
    \end{tabular}
    \label{tab:validation}
\end{table*}


\subsection{Tabular Reporter} Catch2 offers several reporters to format the output such as xml, junit, console, and compact. 
In this work, we implement a TabularReporter class derived from StreamingReporterBase of Catch2 to print all the metrics (mean, standard deviation, and their upper and lower bounds calculated by statistical bootstrapping) in a tabular format.
This reporter can be called at the command line via the arguments `-{}-reporter=tabular' or `-r tabular'.

\subsection{Validation} To validate our framework's output we benchmark the single/double precision generalized matrix multiplication [S/D]GEMM kernels from cuBLAS on Nvidia A6000. 
We compute the average of 100 executions of cuBLAS's SGEMM/DGEMM kernels while measuring the start and end times on the host with std::chrono's clock.
Table \ref{tab:validation} shows that the GFLOPs/sec from the present framework show less than $0.1\%$ deviation from std::chrono's measurements.
This demonstrates that the present framework is capable of measuring kernel's timings accurately with several advantages such as specifying warmup time and statistical bootstrapping.

\subsection{Setup and execution} 
This source code hosted at \url{https://github.com/BNL-HPC/openmp-benchmarks} can be used to generate the data reported in this paper when installed and executed successfully.
Instructions for installation on various machines are listed in the accompanying README.md.
The reproducibility is contingent on access to OLCF resources and allocation on Frontier.
Here, we provide the details for obtaining metrics listed in Table \ref{tab:atomic_capture_frontier} on OLFC Frontier as a table is easier to reproduce than figures.
The branch artifact-repro-frontier contains the compiler flags required for reproducibility on Frontier, however, the correct compiler flags need to be supplied to CMake for different hardware.
The hardware AMD MI250X GPU is available on Frontier nodes whereas the primary software dependency is the OpenMP offload capable compiler. 
While several such compiler versions have been installed in private locations on Frontier, the globally accessible compilers amdclang++ from rocm/5.4.3 and rocm/6.0.0 can be used for  reproducibility. 
Each individual execution for a compiler takes less than a minute with argument -{}-benchmark-samples 100, however, this time increases with the number of samples. 
A successful execution of the artifact will generate the mean and standard deviation of atomic capture for three data types (double, float, int) and two array sizes ($2^{16},2^{20}$) for the selected compiler. 

\section{Results}
\label{sec:results}
The forthcoming examples illustrate the need of a micro-benchmarking framework. 
In all the kernels discussed below, a large array of size $N$ is allocated on the device memory.
It is then initialized with random numbers (RNs) uniformly distributed between $-1$ and $1$ for double and float data types, and between $-100$ and $100$ for integer datatype. 
The benchmark is executed with 1000 samples and 100 resamples with 0.95 confidence interval for statistical bootstrapping \review{and the best performing compiler flags}. 
We compare their performance on Perlmutter (with NVIDIA A100 GPUs), Frontier (with AMD MI250X GPUs), and local servers with Nvidia A6000 and V100 GPUs.

\subsection{Zaxpy Kernel} \label{sec:zaxpy}

\begin{figure*}
\centering
\includegraphics[width=0.8\textwidth]{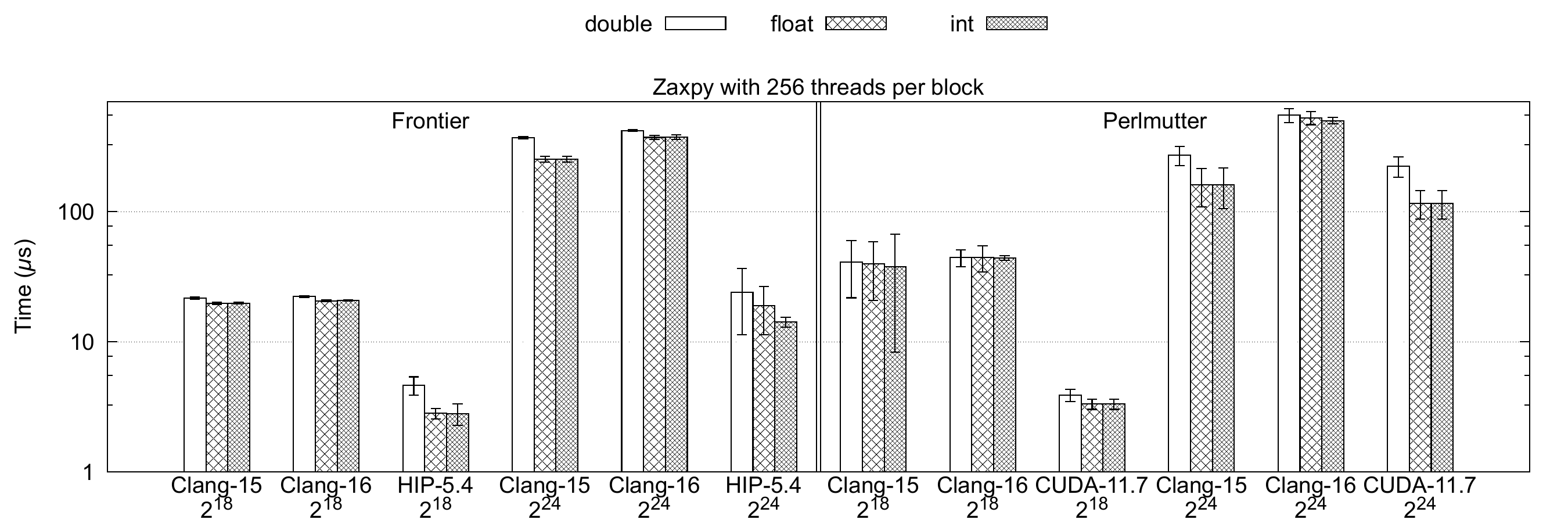}
\caption{Comparison of zaxpy execution time with the native programming models on Frontier and Perlmutter.}
\label{fig:saxpy_perl_2}
\end{figure*}

\begin{figure}
\centering
\includegraphics[width=0.49\textwidth]{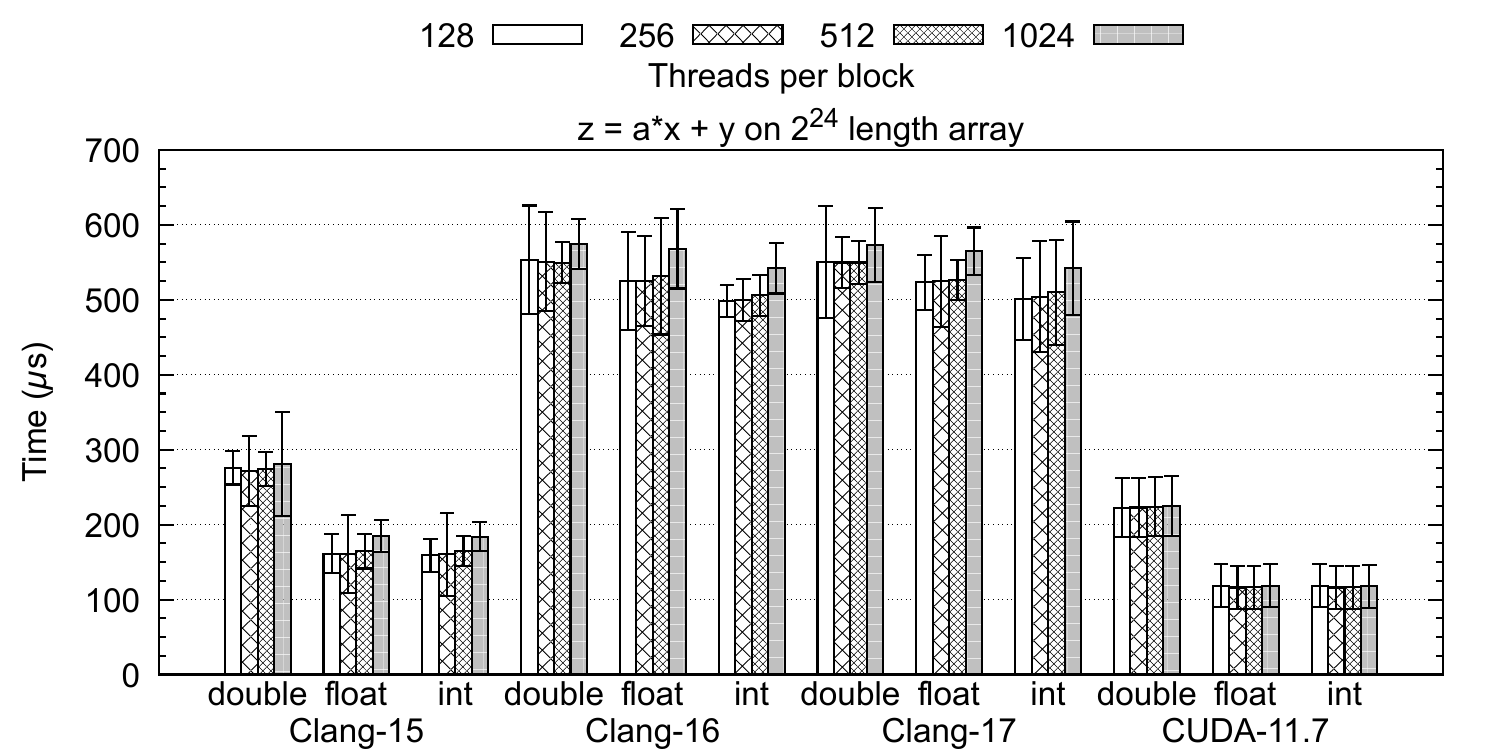}
\caption{Zaxpy execution time on Perlmutter: varying threads per block and datatype.}
\label{fig:saxpy_perl_1}
\end{figure}  

We begin by comparing the performance of the well-known zaxpy kernel which parallelizes the computation of $z=ax+y$, where $x,y,z$ are large arrays and $a$ is a constant.
From Fig. \ref{fig:saxpy_perl_1}, it is seen that for array size $2^{24}$ performance has deteriorated from Clang-15 to Clang-16 and beyond.
The is possibly due to an update in optimization strategy by the compiler which has negatively impacted this particular kernel.
With Clang-15 double datatype was twice as slow as floats and integers whereas for Clang-16 and Clang-17 no such trend is observed.
It can also be seen that the number of threads per block has a marginal impact on the performance.
Similar trend is also observed for an array initialization (see Fig. \ref{fig:array_init_perl}).

Figure \ref{fig:saxpy_perl_2} provides a detailed look into the statistics that the present framework will provide.
Here, we select 256 threads per block and compare performance for two array lengths of $2^{18}$ and $2^{24}$ and report timings on log scale.
The plot compares two versions of LLVM Clang with the respective native programming models on Frontier (HIP) and Perlmutter (CUDA).
It is observed that the previously mentioned deterioration in performance while upgrading to Clang-16 from Clang-15 on Perlmutter is absent for smaller arrays of size $2^{18}$.
On the other hand, on Frontier the performance does not change significantly from Clang-15 to Clang-16.
It is also seen that both native programming models are faster than their OpenMP target offload counterparts.

Although, the plot also enables an overall comparison of the two leadership class machines, it should be noted that (a) the hardware specifications are different for the two machines, and (b) this benchmark is highly simplified whereas most real-world applications are complex with many different types of compute-intensive kernels.


\subsection{\label{sec:atomiccapture}Atomic Capture}

\begin{figure}
\centering
\includegraphics[width=0.45\textwidth]{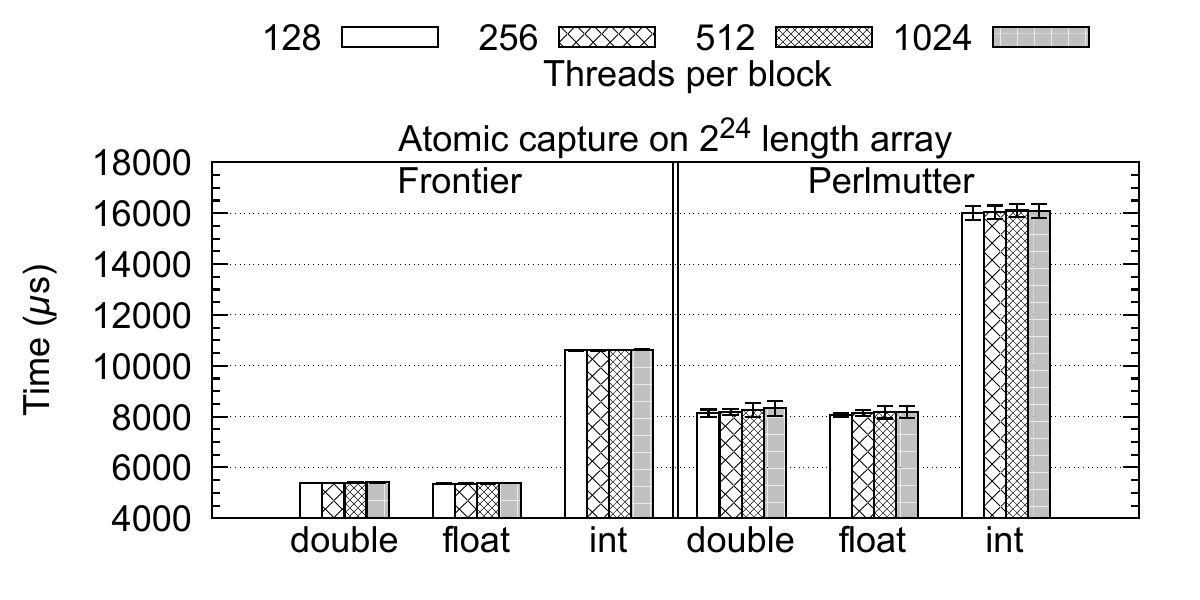}
\caption{Atomic capture on Perlmutter and Frontier with Clang-16: varying threads per block and datatype.}
\label{fig:atomic_capture_per_front}
\end{figure}  

\begin{figure}
\centering
\includegraphics[width=0.45\textwidth]{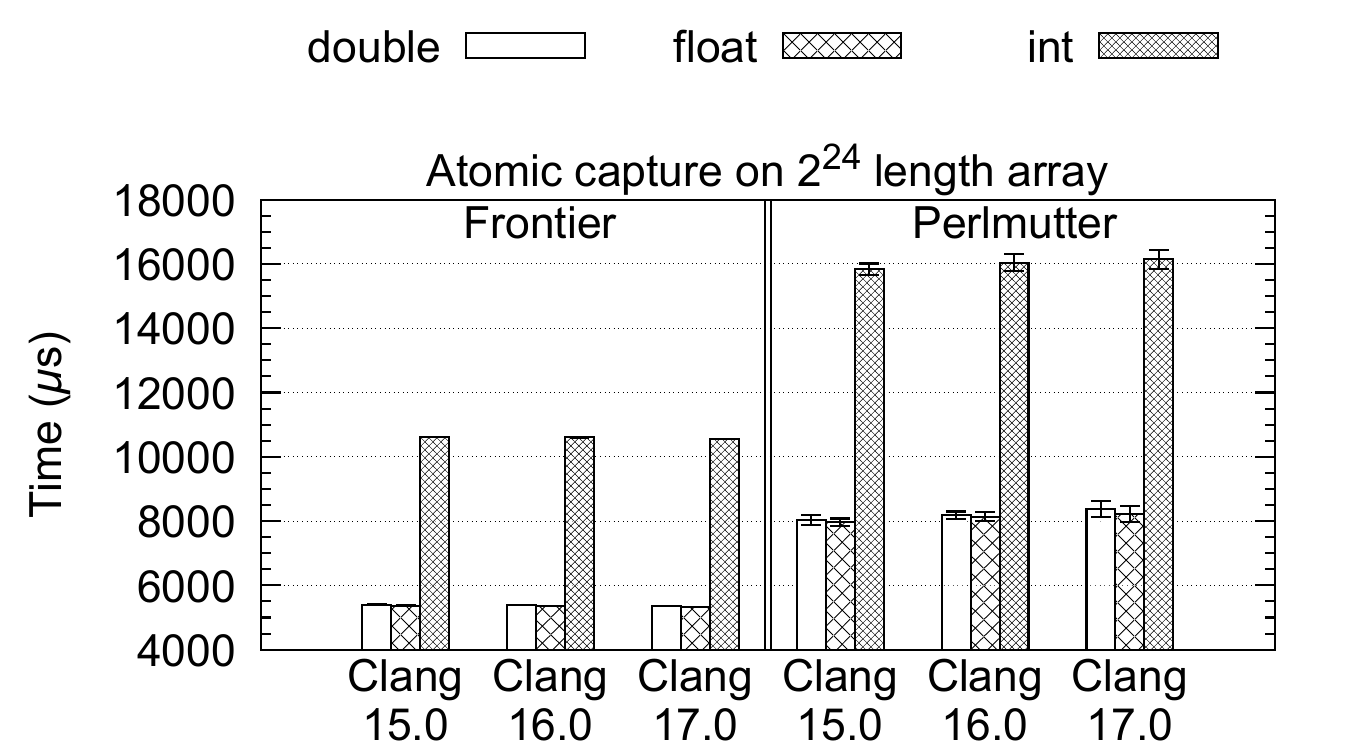}
\caption{Atomic capture on Perlmutter and Frontier with 256 threads per block and different datatype.}
\label{fig:atomic_capture_per_front_256}
\end{figure}  

\begin{figure*}
\centering
\includegraphics[width=0.75\textwidth]{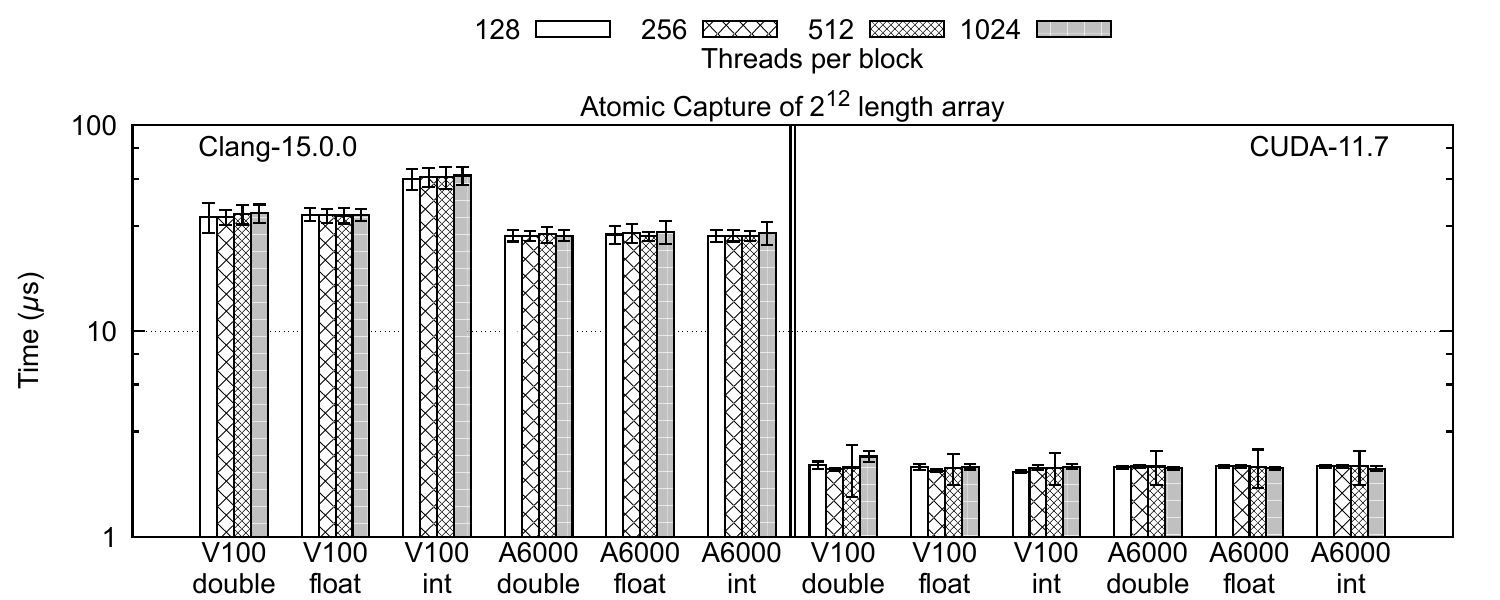}
\includegraphics[width=0.77\textwidth]{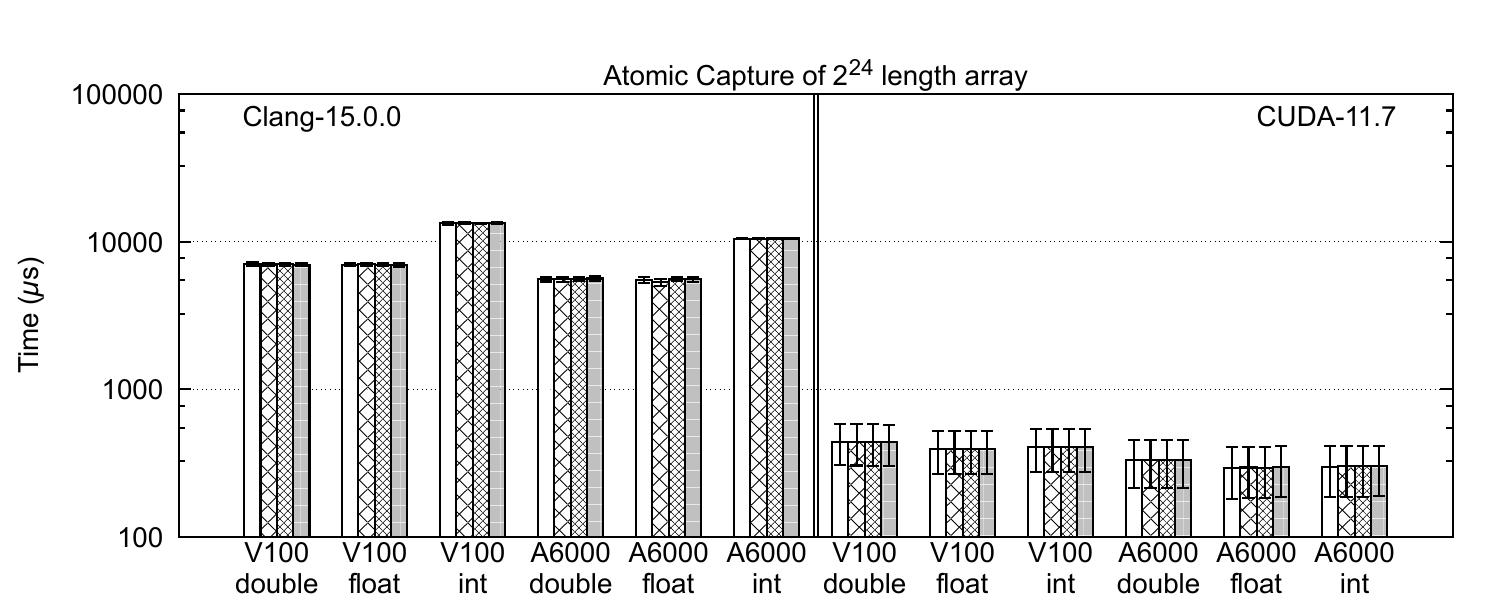}
\caption{Comparison of OpenMP's atomic capture construct on Nvidia V100 and A6000 with CUDA. }
\label{fig:atomic_capture_nvidia_v100_a6000}
\end{figure*}  

In this section, we use OpenMP's \verb|atomic capture| construct to develop a benchmark for capturing positive members of a large array.
The objective is to collect only the positive RNs in contiguous memory locations of an array while also keeping a count of the total number of positive RNs.
This operation is encountered frequently in several scientific applications that require tracking two indices \cite{atif2024porting}.
A simple algorithm for executing this operation with multiple threads and employs \verb|atomic capture| construct of OpenMP.

Figure \ref{fig:atomic_capture_per_front} compares the performance of \verb|atomic capture| on Frontier and Perlmutter using different numbers of threads per blocks and data types with Clang-16. 
It can be seen that this operation is faster on Frontier for all data types. 
Unlike \verb|zaxpy| and array initialization benchmark, we do not observe an appreciable variation with number of threads.
Figure \ref{fig:atomic_capture_per_front_256} compares performance of three recent versions of LLVM Clang, all of which perform similarly.
It can be seen that Perlmutter has larger standard deviation for 1000 samples considered here for arrays of size $2^{24}$, however, similar trends were observed for arrays of size $2^{20}$ and $2^{28}$.
It should also be noted that integer performs poorly in comparison to double/float data types on both machines.

In order to compare the performance of OpenMP's atomic construct with a native programming model, we plot the timings on Nvidia A6000 and V100 in Fig. \ref{fig:atomic_capture_nvidia_v100_a6000}. 
It can be seen that the native programming model CUDA is much faster than Clang-15 for both small and large array sizes.
The number of threads per block has marginal impact on the metrics, however, made exception for small arrays on A6000, integer data types are slower than double and float for Clang whereas CUDA's performance is uniform across all data types.



\subsection{\label{sec:atomicupdate}Atomic Update}

\begin{figure}
\centering
\includegraphics[width=0.49\textwidth]{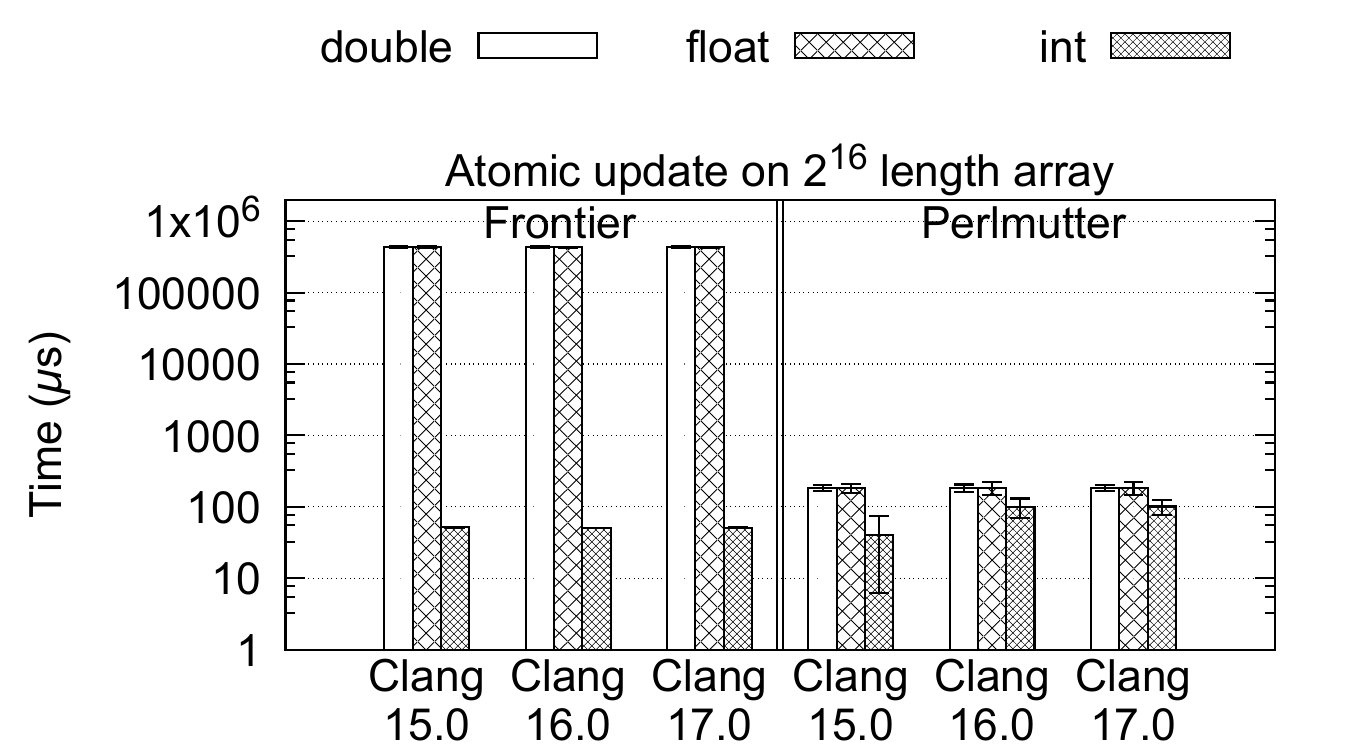}
\caption{Atomic update on Perlmutter and Frontier with 256 threads per block and different datatype.}
\label{fig:atomic_update_per_front_256}
\end{figure}  

\begin{figure}
\centering
\includegraphics[width=0.47\textwidth]{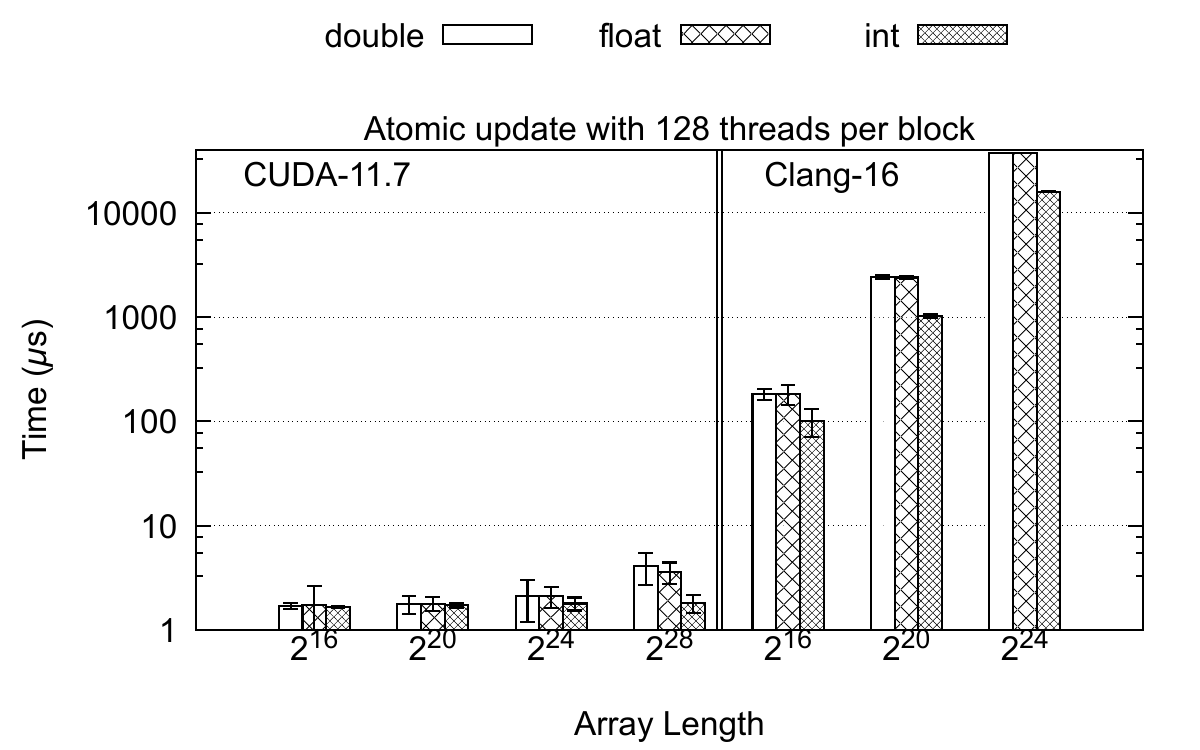}
\caption{Atomic update on Perlmutter: Comparison of CUDA and Clang-16.}
\label{fig:atomic_update_perl_128_cuda_vs_clang}
\end{figure}  

\begin{figure*}
\centering
\includegraphics[width=0.75\textwidth]{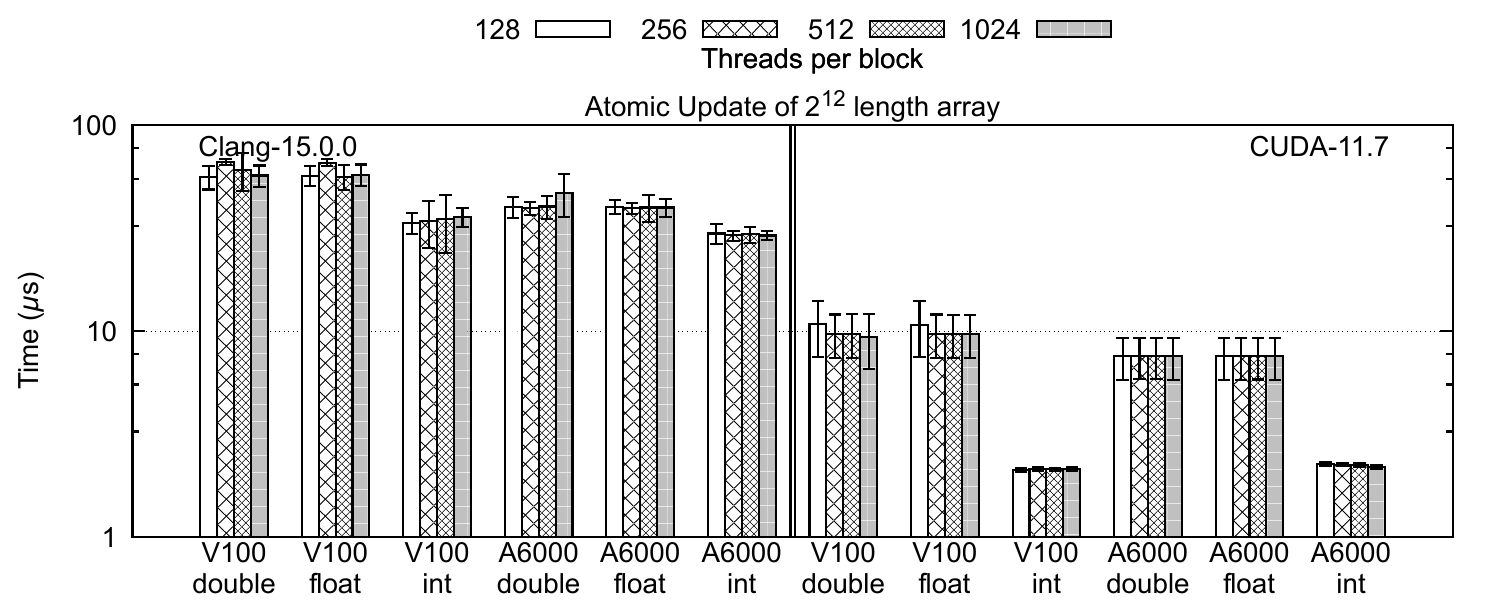}
\includegraphics[width=0.75\textwidth]{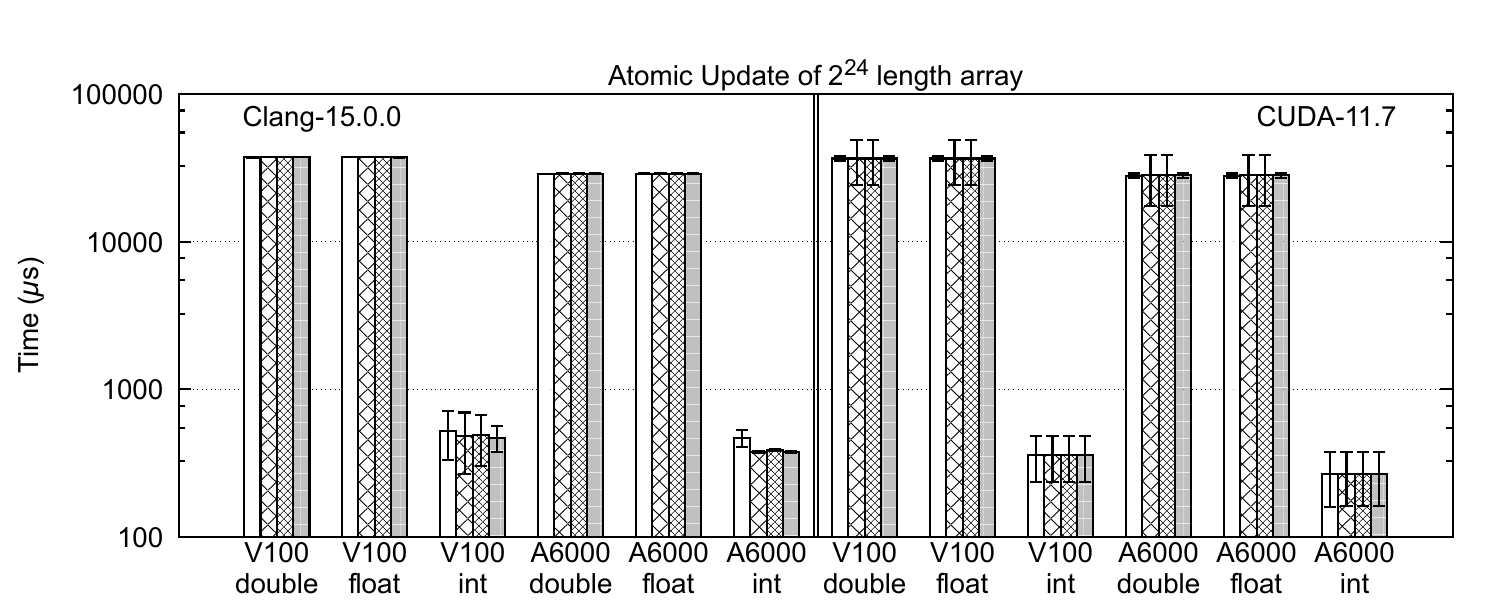}
\caption{Comparing atomic update time: varying threads per block and datatype double and int.}
\label{fig:atomic_update_a6000_v100}
\end{figure*}

\begin{figure*}
\centering
\includegraphics[width=0.8\textwidth]{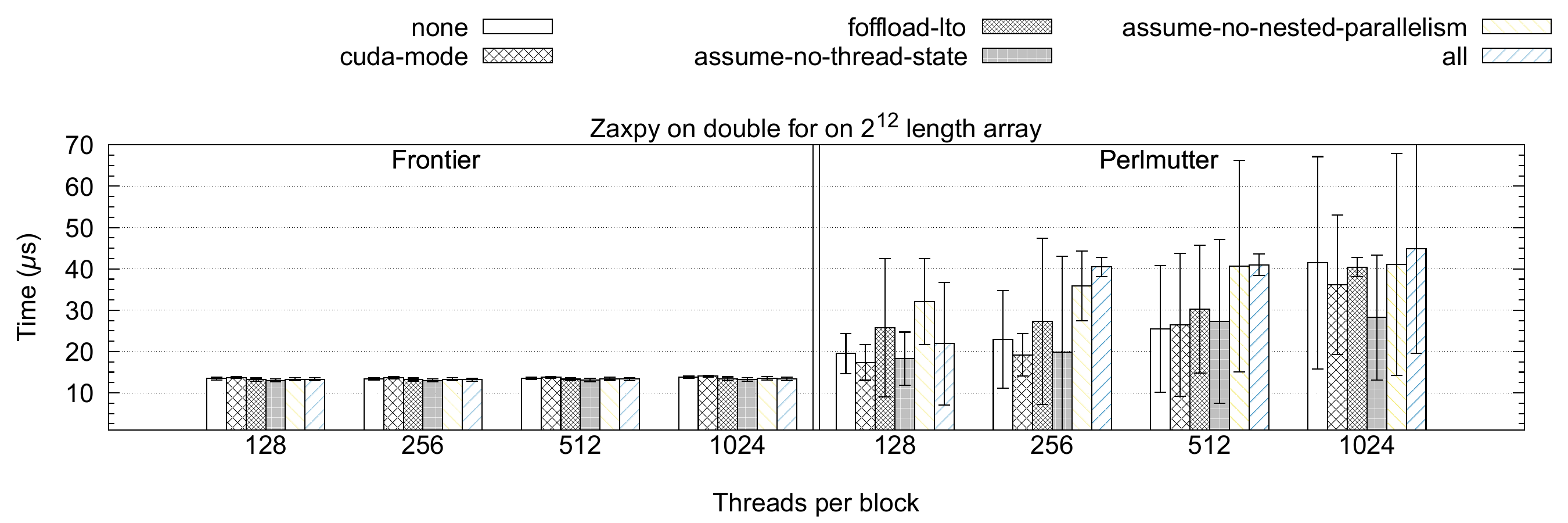}
\caption{Evaluating the impact of LLVM Clang's compiler flags on zaxpy kernel }
\label{fig:zaxpy_compiler_flags_212}
\end{figure*}  

\begin{figure}
\centering
\includegraphics[width=0.49\textwidth]{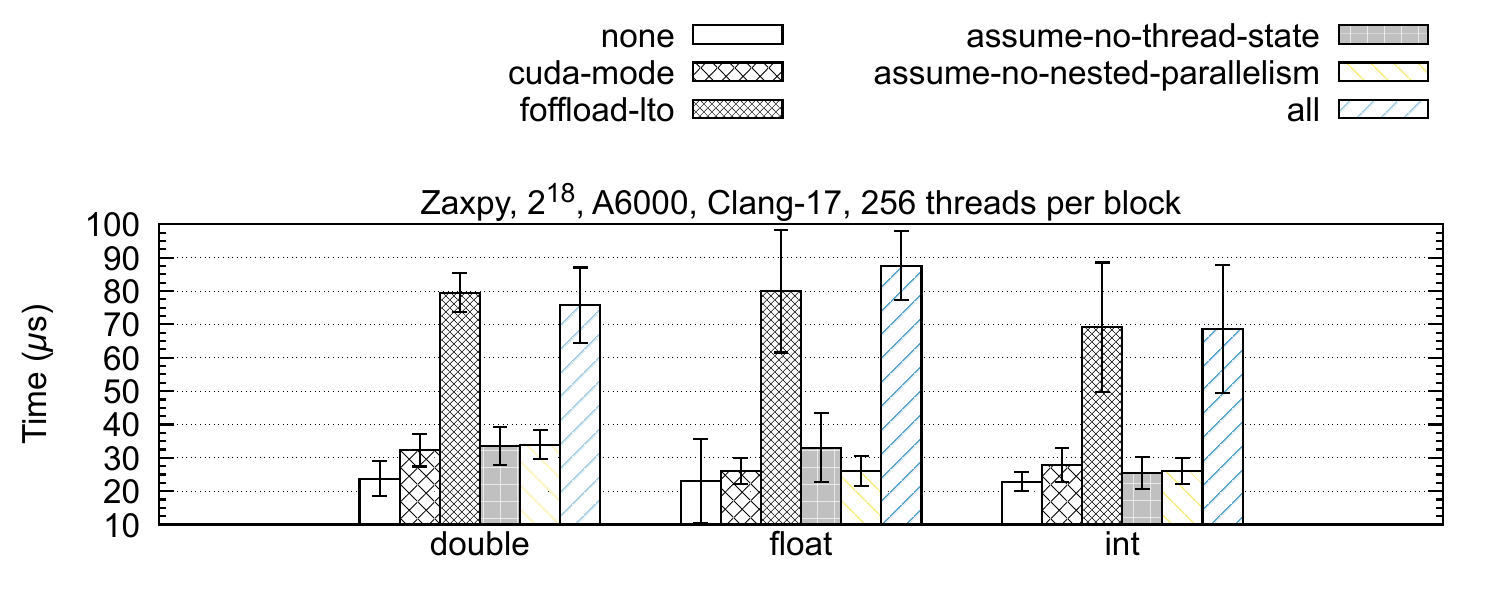}
\includegraphics[width=0.49\textwidth]{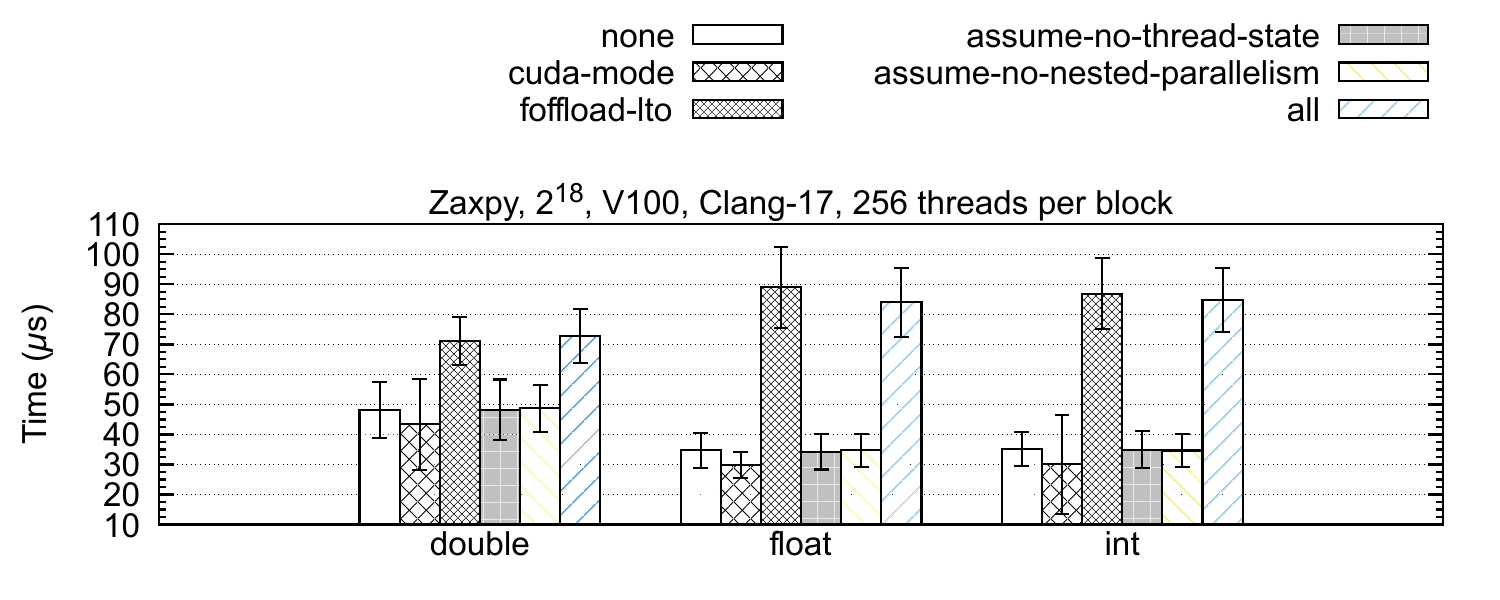}
\caption{Impact of LLVM Clang compiler flags on Nvidia V100 and A6000.}
\label{fig:compiler_flags_clang17}
\end{figure} 

In this benchmark, we use OpenMP's  \verb|atomic update| construct to sum all the elements of a large array. 
Although this operation in practice  performs better as a parallel reduction, here we want to compare the raw performance across LLVM Clang version and hardware.
From Fig \ref{fig:atomic_update_per_front_256} it can be seen that for a relatively small array ($2^{16} = 65,536$ elements), Frontier and Perlmutter yield similar performance for integer datatype. 
The loss in performance for integers is also observed when upgrading from Clang-15 to Clang-16 on Perlmutter, but not on Frontier. 
However, for doubles and floats Frontier turns out to be much slower than Perlmutter. 
Similar performance degradation on AMD platforms have also been observed in ATLAS FastCaloSim \cite{atif2024porting,atif2023evaluating}.

Fig \ref{fig:atomic_update_perl_128_cuda_vs_clang} compares timings for \verb|atomic update| on Perlmutter for different array sizes. 
It is seen that CUDA exhibits similar performance for array sizes $2^{16},2^{20},2^{24}$ for all datatypes, however, for Clang-16 integer data types are faster compared to doubles and floats.
We also observe that going from $2^{16},2^{20},2^{24}$ to $2^{28}$ the timing increases by a factor of two for CUDA with doubles and floats, however, the they grow exponentially for Clang-16 even for smaller array sizes.
LLVM Clang-16, on the other hand, is about $75\times$ slower for $2^{16}$ sized array and the timings grow exponentially with array size.

Figure \ref{fig:atomic_update_a6000_v100} plots the timings \verb|atomic update| across data types and threads per block for Nvidia V100 and A6000.
It stands out that CUDA performs much better than Clang for a smaller array size of $2^{12}$, however, this difference diminishes as the array sizes are increased to $2^{18}$ and $2^{24}$.
It is also seen that smaller array sizes have smaller disparity between the performance of integers and floats whereas the difference becomes more prominent for larger array sizes.
We also note that doubles and floats exhibit similar performance in all the test runs.
\review{We also note that integers perform better that floats in atomic update but worse in atomic capture.
This could likely be attributed to to specialize-hardware support. 
An atomic capture combines the update with a read of the previous value in a single operation which being a more complex operation may not have specialized hardware support. 
However, further investigation is required to support the above hypothesis.}


\subsection{Impact of Compiler Flags}

In the previous examples, all the available compiler flags listed in Sec. \ref{sec:openmp} were enabled. 
Severe performance deterioration has been observed on some GPUs when the correct set of flags were not used.
For examples, in ATLAS FastCaloSim we observe a performance gain of 10x upon using \verb|-fopenmp-cuda-mode| flag on Nvidia GPUs but not on AMD GPUs.
However, there are no comprehensive studies demonstrating the impact of compiler flags on minimal working examples.
With the present framework, it is straightforward to test if a compiler flag has been introduced that will improve performance. 
Thus in this section we briefly assess the impact of different compiler flags using Clang-16.
From Fig. \ref{fig:zaxpy_compiler_flags_212}, it is seen that the change in performance upon using different compiler flags is observed for Perlmutter, whereas their impact is negligible on Frontier.
The standard deviation on Perlmutter is large, but it is seen that the cuda-mode and assume-no-thread-state flags perform better than the foffload-lto flag for the zaxpy kernel. 
The figure also illustrates the need to investigate the observed large standard deviations on Perlmutter.


The impact of compiler flags is more dominant on Nvidia A6000 and V100 as shown in Fig. \ref{fig:compiler_flags_clang17}.
It is seen that \verb|offload-lto| flag leads to a larger loss in performance.
\review{In principle, the link-time optimizations must improve the performance of GPU kernels. However due to the linear nature of of zaxpy it might be well optimized without the flags and the} \verb|offload-lto| \review{flag may have conflicted with the base optimizations.}
On A6000, using no compiler flags is consistently the fastest for all data types, however, \verb|-fopenmp-cuda-mode| slightly improves performance on V100 for all data types.
The \verb|-fopenmp-assume-no-threadstate| and \verb|-fopenmp-no-nested-parallelism| flags have marginal impact on performance of the zaxpy kernel.
In this section, we have demonstrated an example of how choosing an incompatible compiler flag can diminish performance for a simple zaxpy kernel, however, the framework can be used to analyse performance impacts on many complex kernels across different GPUs too.

\begin{table}[]
    \centering
    \caption{Frontier: Atomic Capture, 256 threads per block, mean (std) of execution times in $10^{-6}$secs}
    \begin{tabular}{c|c|c|c}
    \midrule
    $2^{16}$ array & double & float & int \\
    \midrule
    Clang15 & 44.27 (0.95) & 44.11 (1.14) & 64.85 (1.05) \\
    Clang16 & 39.13 (0.37) & 39.16 (0.37) & 60.28 (0.49) \\
    Clang17 & 39.07 (0.47) & 39.05 (0.49) & 59.91 (0.57) \\
    Clang18 & 39.87 (0.47) & 40.38 (0.45) & 39.52 (0.47) \\
    Clang19 & 40.71 (0.51) & 40.01 (0.46) & 40.23 (0.53) \\
    Clang20 & 40.71 (0.37) & 39.80 (0.51) & 40.45 (0.49) \\
    rocm543 & 47.27 (1.02) & 46.95 (0.97) & 46.55 (1.14) \\
    rocm600 & 34.29 (0.27) & 34.01 (0.28) & 34.19 (0.39) \\
    afar2512 & 22.49 (0.28) & 22.13 (0.35)  & 22.16 (0.36) \\
    \midrule
    $2^{20}$ array & double & float & int \\
    \midrule
    Clang15 & 364.25 (3.22) & 362.11 (3.12) & 699.49 (2.89) \\
    Clang16 & 358.57 (3.02) & 357.48 (3.35) & 697.95 (2.16) \\
    Clang17 & 347.11 (0.45) & 345.80 (0.51) & 671.82 (1.17) \\
    Clang18 & 351.86 (0.63) & 348.86 (0.69) & 345.46 (0.43) \\
    Clang19 & 352.13 (0.57) & 349.03 (0.52) & 346.06 (0.52) \\
    Clang20 & 352.10 (1.93) & 349.05 (0.47) & 345.76 (0.60) \\
    rocm543 & 377.44 (8.21) & 374.98 (7.96) & 372.25 (8.06) \\
    rocm600 & 49.30 (1.85) & 47.85 (0.23) & 47.88 (0.20) \\
    afar2512 & 32.58 (0.53) & 30.33 (0.43) & 30.40 (0.49) \\
    \midrule
    \end{tabular}
    \label{tab:atomic_capture_frontier}
\end{table}

\subsection{Compilers and compiler version}

As the compilers supporting OpenMP's target offload model are undergoing rapid developments we have observed dramatic change in performance across GPUs.
The present framework can be used by developers to select the best compiler and version for their applications.
It was previously reported in Fig. \ref{fig:saxpy_perl_1} that the performance of the zaxpy kernel had deteriorated after upgrading the Clang version.
We list the timings for \verb|atomic capture| on Frontier obtained from different compilers and their version  in Table \ref{tab:atomic_capture_frontier}.
It is observed that even though the mean does not change by Clang upgrades, the standard deviation has reduced as the compiler version was upgraded for both array sizes $2^{16}$ and $2^{20}$.
The table also demonstrates the gain in performance obtained by simply upgrading rocm-5.4.3 to rocm-6.0.0 or using a different compiler such as AMD's AFAR.

\section{Discussion and Outlook}
\label{sec:outlook}

We have presented a framework for developing micro-benchmarks with the aim of comparing OpenMP's \verb|target offload| model with the native programming models.
This framework simplifies tracking the evolution of various compilers.
We have discussed a few commonly used operations and their performance, however, incorporating more benchmarks into the framework is straightforward.
The \verb|Catch2| library used here computes the metrics via statistical sampling and bootstrapping. 
Thus, the performance numbers generated here are different from running a profiler on a loop of the benchmarked kernels.
This is because \verb|Catch2| incorporates a smart way of avoiding the case of the compiler optimizing away the same kernel executed in a loop.
The statistical sampling also reduces the chance of a fringe case biasing the metrics.
As \verb|Catch2| is a testing framework, the benchmarks also include assert conditions that ensure correctness and give insight into precision loss.
These benchmarks can also serve as minimal working examples to demonstrate performance changes to compiler developers.

In the future, we will extend our list of benchmarks to include more atomic and other commonly used operations such as general matrix and vector operations. 
We also plan to compare performance of other vendor and community developed compilers.
In the current benchmarks we have used device memory allocation APIs, but we will also extend examples for using managed memory.
We also plan to include a SYCL port of benchmarks to compare performance on the Intel GPUs. 
\review{Another challenging but critical topic for future is to split the measure times into launch, execution, synchronization without instrumenting the code.}

\section*{Acknowledgment}

This research used resources of the NERSC Center, a U.S. Department of Energy (DOE) Office of Science User Facility located at Lawrence Berkeley National Laboratory, operated under Contract No. DE-AC02-05CH11231,
and of the OLCF at the Oak Ridge National Laboratory, which is supported by the Office of Science of the U.S. DOE under Contract No. DE-AC05-00OR22725.






\bibliographystyle{IEEEtran}
\bibliography{omptargetatomic}

\end{document}